\documentclass{PoS}
\usepackage{amsmath}

\title{Two-Higgs-doublet model fits with HEPfit}

\ShortTitle{2HDM fits with HEPfit}

\author{\speaker{Otto Eberhardt}\\
        Instituto de F\'{i}sica Corpuscular, Universitat de Val\`{e}ncia -- Consejo Superior de
        Investigaciones Cient\'{i}ficas, Parc Cient\'{i}fic, E-46980 Paterna, Valencia, Spain\\
        E-mail: \email{otto.eberhardt@ific.uv.es}}

\abstract{The Two-Higgs-doublet model (2HDM) is one of the most studied extensions of the Standard Model. Like the other popular ``New Physics'' models, it gets more and more constrained by recent experimental progress, especially by the LHC data. For the 2HDM types I and II with a softly broken $\mathbb{Z}_2$ symmetry, I present updated results of global analyses obtained with the open-source HEPfit code, emphasizing the impact of the pre-EPS-HEP 2017 LHC data. Furthermore, I discuss the status of implementation into HEPfit of 2HDM's beyond the conventional $\mathbb{Z}_2$ symmetric types.}

\FullConference{The European Physical Society Conference on High Energy Physics\\
		 5-12 July\\
		 Venice, Italy}

\begin{document}

\section{Introduction}

In spite of no direct indication for physics beyond the Standard Model of particle physics at the LHC, there are many good experimental and theoretical reasons to assume that the Standard Model (SM) is an incomplete description of our world. Lacking distinct measurements of new phenomena, two things become crucial for the analyses of SM extensions: The precise understanding of the SM as well as the consistent combination of all information on possible indirect signs of ``New Physics'' we can gather. In the following, I will present such an analysis for one of the most popular models that extend the SM, the Two-Higgs-Doublet model. It adds a second Higgs doublet to the SM particle content. In the past, many different constraints on this model have been explored; recently, the discovery of the 125 GeV scalar \cite{Aad:2012tfa,Chatrchyan:2012xdj} as well as the absence of a second scalar resonance at the LHC have put strong bounds on the existence of a second Higgs doublet. Here, I will quantify these bounds in the light of updated LHC data, performing global fits to the Two-Higgs-Doublet models with a softly broken $\mathbb{Z}_2$ symmetry (2HDM) of type I and II. Before going into detail, I will give an introduction to the fitting framework HEPfit, which also guarantees the consistent treatment of the SM part at the best precision available.

\section{HEPfit}

As statistical setup for the global 2HDM fits I use the open-source C++ code HEPfit \cite{hepfit}, which is linked to the Bayesian Analysis Toolkit (BAT) \cite{Caldwell:2008fw}. HEPfit calculates flavour and Higgs observables as well as electroweak Z-pole observables, most of them at the best known precision. It can be linked to other programs as a library, but it also comes with an interface to BAT and can be used to perform global fits in the SM and several of its extensions (see also the other HEPfit contributions at EPS-HEP 2017 \cite{EPSValli,EPSSilvestrini,EPSPaul}). The release of the first fully documented HEPfit version is planned in the near future.

\section{The 2HDM}

The most general formulation of Two-Higgs-Doublet models \cite{Lee:1973iz,Gunion:2002zf,Branco:2011iw} is characterized by the Higgs potential
\begin{align*}
V_H^{\text{\tiny{2HDM}}}=
& \; m^2_{11}\Phi_1^\dagger \Phi_1
+m^2_{22}\Phi_2^\dagger \Phi_2^{\phantom{\dagger}}
-\left( m_{12}^2 \Phi_1^\dagger \Phi_2^{\phantom{\dagger}}
+{\rm H.c.} \right) \\
& \; +\frac{\lambda_1}{2} \left( \Phi^\dagger _1\Phi^{\phantom{\dagger}}_1 \right) ^2
+\frac{\lambda_2}{2} \left( \Phi_2^{\dagger}\Phi_2^{\phantom{\dagger}} \right) ^2
+\lambda_3 \left( \Phi_1^{\dagger}\Phi_1^{\phantom{\dagger}}\right)  \left( \Phi_2^{\dagger}\Phi_2^{\phantom{\dagger}}\right)
+\lambda_4 \left( \Phi_1^{\dagger}\Phi_2^{\phantom{\dagger}}\right)  \left( \Phi_2^{\dagger}\Phi_1^{\phantom{\dagger}}\right) \\
& \; +\left[ \frac{\lambda_5}{2} \left( \Phi_1^{\dagger}\Phi_2^{\phantom{\dagger}}\right) ^2
+\lambda_6 \left( \Phi_1^{\dagger}\Phi_1^{\phantom{\dagger}}\right) \left( \Phi_1^{\dagger}\Phi_2^{\phantom{\dagger}}\right)
+\lambda_7 \left( \Phi_2^{\dagger}\Phi_2^{\phantom{\dagger}}\right) \left( \Phi_1^{\dagger}\Phi_2^{\phantom{\dagger}}\right)
+{\rm H.c.} \right] ,
\end{align*}
where $\Phi_1$ and $\Phi_2$ are the two Higgs doublets.\\
The corresponding Yukawa Lagrangian reads

\begin{align*}
  {\cal L}_{\text{\tiny{Yukawa}}} &= -\sum\limits_{j,k=1}^3 \Big[ Y^{d,1}_{jk}\left( \bar{Q}_j \Phi_1 \right) d_k +Y^{d,2}_{jk}\left( \bar{Q}_j \Phi_2 \right) d_k \Big. \\[-8pt]
  &\hspace*{50pt}+Y^{u,1}_{jk} \left( \bar{Q} _j{\rm i}\sigma _2 \Phi^*_1 \right) u_k +Y^{u,2}_{jk} \left( \bar{Q} _j{\rm i}\sigma _2 \Phi^*_2 \right) u_k\\[2pt]
  &\hspace*{50pt}\Big.+Y^{\ell,1}_{jk} \left( \bar{L}_j \Phi_1 \right) \ell _k +Y^{\ell,2}_{jk} \left( \bar{L}_j \Phi_2 \right) \ell _k+{\rm H.c.}\Big] ,
\end{align*}
with the left-handed fermion fields $Q$ and $L$ and the right-handed fermion fields $u$, $d$ and $\ell$.

While the HEPfit collaboration is working at an implementation of the most general model into HEPfit, I will focus here on the case without explicitly broken $\mathbb{Z}_2$ symmetry. This implies $\lambda_6=\lambda_7=Y^{u,1}=0$ and either $Y^{d,1}=Y^{\ell,1}=0$ (type I) or $Y^{d,2}=Y^{\ell,2}=0$ (type II).\footnote{For the cases with $Y^{d,1}=Y^{\ell,2}=0$ or $Y^{d,2}=Y^{\ell,1}=0$ I refer to \cite{Chowdhury:2017xxx}.} For an example of a 2HDM fit to a more general Yukawa sector with HEPfit, see \cite{EPSPaul}.
I furthermore assume that all couplings in $V_H^{\text{\tiny{2HDM}}}$ are real and that the 125 GeV scalar is the lightest 2HDM scalar $h$. The other physical Higgs particles are the neutral scalar $H$, the neutral pseudoscalar $A$ and the charged scalars $H^\pm$. In the fits, I assume that their masses as well as the soft $\mathbb{Z}_2$ breaking scale $|m_{12}|$ are below 1.5 TeV.
Apart from these masses, the 2HDM is defined by the two mixing angles $\alpha$ and $\beta$ between these scalars, instead of which I will use $\tan \beta$ and $\beta-\alpha$. The SM parameters are fixed to their best fit values \cite{deBlas:2016ojx}.

\section{Constraints}

As mentioned before, I want to emphasize the impact of the LHC measurements on the 2HDM parameter space. They can be divided into the signal strengths of the 125 GeV resonance $h$ and the searches for heavier scalars. For both, I use all available data from the 7+8 TeV run and the 13 TeV run which was made public before the EPS-HEP 2017 conference: signal strengths of $h$ decaying to $\gamma \gamma$, $bb$, $\tau\tau$, $\mu\mu$, $WW$ and $ZZ$ \cite{Khachatryan:2016vau,ATLAS-CONF-2016-063,ATLAS-CONF-2016-080,ATLAS-CONF-2016-081,ATLAS-CONF-2016-091,ATLAS-CONF-2016-112,CMS-PAS-HIG-16-003,CMS-PAS-HIG-16-020,CMS-PAS-HIG-16-033,CMS-PAS-HIG-16-043,CMS-PAS-HIG-16-038,CMS-PAS-HIG-17-003} and the search for heavy neutral resonances in decays to $bb$, $\tau\tau$, $\gamma \gamma$, $Z\gamma$, $ZZ$, $WW$, $hh$ and $hZ$ \cite{Aad:2014fha,Aad:2014ioa,Aad:2014vgg,Aad:2015agg,Aad:2015kna,Aad:2015wra,Aad:2015xja,ATLAS-CONF-2016-056,ATLAS-CONF-2016-062,ATLAS-CONF-2016-059,ATLAS-CONF-2016-085,ATLAS-CONF-2016-074,ATLAS-CONF-2016-044,ATLAS-CONF-2016-082,ATLAS-CONF-2016-079,ATLAS-CONF-2016-071,ATLAS-CONF-2016-004,ATLAS-CONF-2016-015,ATLAS-CONF-2016-017,Khachatryan:2015cwa,Khachatryan:2015lba,Khachatryan:2015tha,Khachatryan:2015tra,Khachatryan:2015yea,Khachatryan:2016sey,CMS-PAS-HIG-14-029,CMS-PAS-HIG-16-014,CMS-PAS-EXO-16-035,CMS-PAS-EXO-16-027,CMS-PAS-HIG-16-033,CMS-PAS-HIG-16-023,CMS-PAS-HIG-16-029,CMS-PAS-HIG-16-025,CMS-PAS-EXO-16-034,CMS-PAS-HIG-16-011,CMS-PAS-HIG-16-037,CMS-PAS-HIG-16-002,CMS-PAS-HIG-16-032,CMS-PAS-HIG-15-013,CMS-PAS-HIG-16-034} as well as the search for charged scalars with the final states $\tau \nu$ and $tb$ \cite{Aad:2014kga,Khachatryan:2015qxa,Aad:2015typ,ATLAS-CONF-2016-088,CMS-PAS-HIG-16-031,ATLAS-CONF-2016-089,ATLAS-CONF-2016-104}. The details of the implementation of these observables into HEPfit can be found in \cite{Cacchio:2016qyh,Chowdhury:2017xxx}; here, I assume that the observed upper limits on the cross sections are identical with the expected ones, given that almost everywhere the two are compatible at the $2\sigma$ level.

On top of the LHC data, I apply a conservative choice for the theoretical constraints: In order to require that the vacuum is stable, I need to guarantee that $V_H^{\text{\tiny{2HDM}}}$ is bounded from below \cite{Deshpande:1977rw} and that the electroweak minimum is the global minimum \cite{Barroso:2013awa}. Moreover, I demand that the eigenvalues of the scattering matrix of two-scalar-to-two-scalar scattering processes do not exceed 1 in magnitude \cite{Ginzburg:2005dt}, and that the next-to-leading order contribution to these eigenvalues is not larger than its leading order value \cite{Grinstein:2015rtl,Cacchio:2016qyh}.
Finally, I combine the mentioned bounds with the remaining relevant constraints: The 2HDM should be in agreement with electroweak precision data, so I use the latest HEPfit values \cite{deBlas:2016ojx,EPSSilvestrini} for the Peskin-Takeuchi pseudo-observables $S$, $T$ and $U$ \cite{Peskin:1990zt,Peskin:1991sw,Haber:1993wf}. I also include the two most relevant flavour observables to the fit, namely ${\cal B}(b\to s\gamma)$ and $\Delta m_{B_s}$ \cite{Misiak:2015xwa,Geng:1988bq,Deschamps:2009rh,Amhis:2016xyh}.

\section{Fit results}

\begin{figure}
\begin{picture}(450,130)(0,0)
\put(-10,0){\includegraphics[width=450pt]{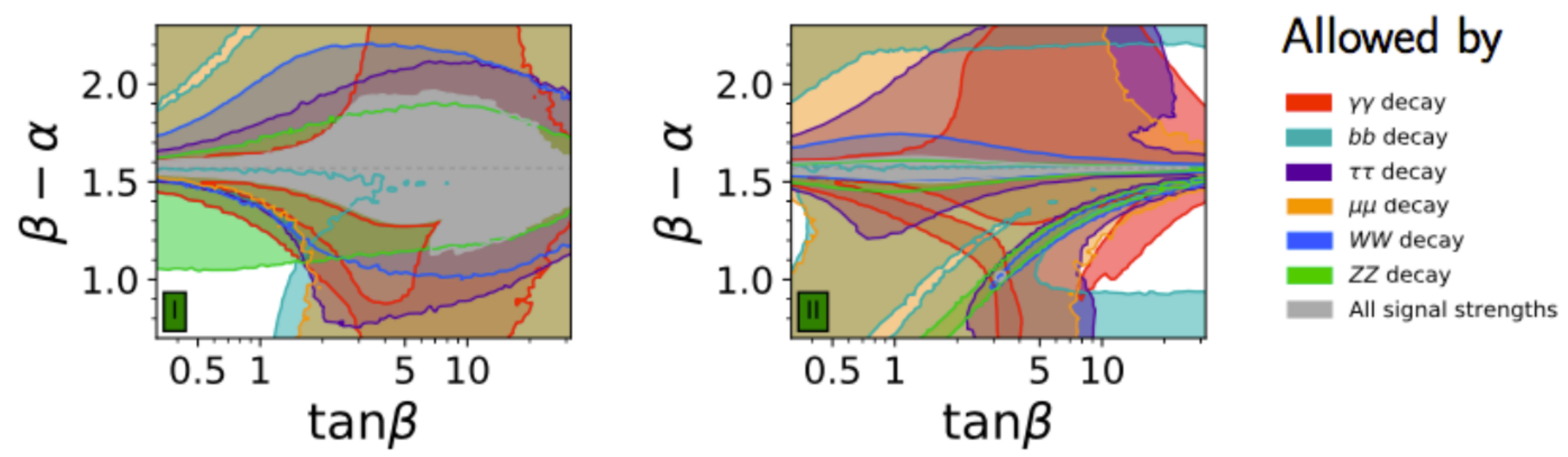}}
\end{picture}
\caption{$95$\% allowed regions in the $\beta-\alpha$ vs. $\tan \beta$ plane. The coloured contours are allowed by the $h$ decays to $\gamma \gamma$, $bb$, $\tau\tau$, $\mu\mu$, $WW$ and $ZZ$, respectively (see the legend). The grey area results from the combination of all signal strengths. In the left panel, the type I fit result is displayed and in the right panel the one from the type II fit.}
\label{fig:1}
\end{figure}

To start with, I discuss the effect the $h$ signal strengths have on the 2HDM parameters. Since the tree-level couplings of $h$ to fermions and gauge bosons only depend on the 2HDM angles, it is obvious to study the $\beta-\alpha$ vs.~$\tan \beta$ plane. For both types, these planes are shown in Figure \ref{fig:1} with the single contributions of all relevant $h$ decays as well as their combination. While $\tan\beta$ can have any value between 0.3 and 30, the difference between $\beta$ and $\alpha$ is forced to be close to the so-called alignment limit of $\pi/2$ for which the $h$ couplings become SM-like. The maximal deviation of this value depends on $\tan\beta$ and the 2HDM type and is smaller compared to the fits to data from the 7+8 TeV run of the LHC (see e.g. \cite{EPSPeiffer,Chowdhury:2015yja}). In the low $\tan\beta$ range, the $\gamma \gamma$ signal strengths prevail in type I, whereas for $\tan\beta>8$ the $h\to ZZ$ measurements are the strongest. In type II, the most stringent bounds come from the $ZZ$ and $WW$ signal strengths. The tree-level coupling of the $h$ boson to massive gauge bosons is type independent, but these constraints also depend on the fermion couplings in the $h$ production and decay width; that is why the bounds are much stronger in type II. For the latter, it is also worth mentioning that the so-called wrong-sign Yukawa coupling solution is only compatible with all signal strengths in a very small region around $\tan\beta=3.2$ and $\beta-\alpha=1$.

\begin{figure}
\begin{picture}(450,220)(0,0)
\put(0,0){\includegraphics[width=360pt,trim=0 420 10 10,clip=true]{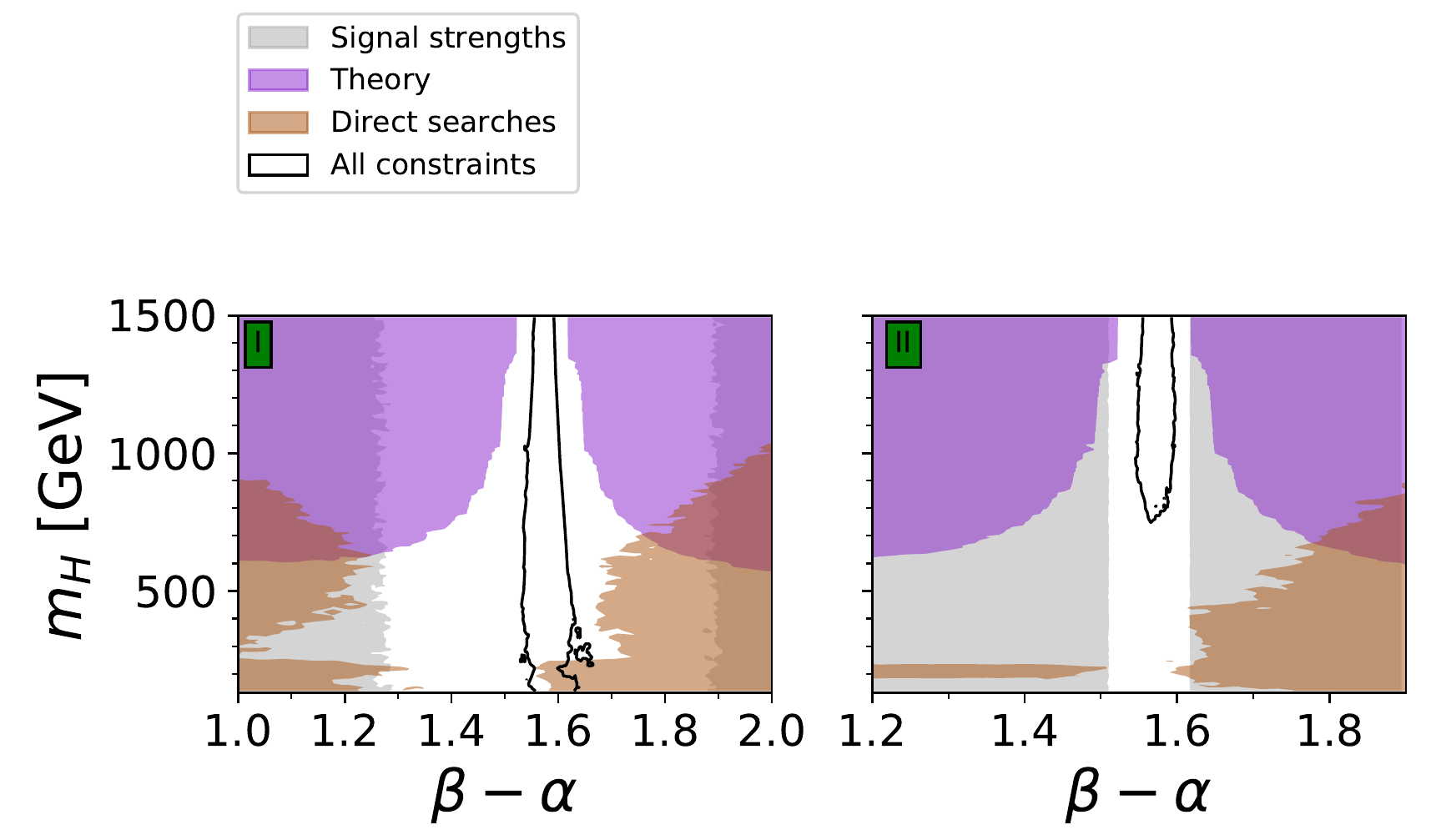}}
\end{picture}
\caption{$95$\% excluded regions in the $m_H$ vs. $\beta-\alpha$ planes of type I (left) and II (right). The exclusions stem from all signal strengths (grey), all heavy Higgs searches (orange), theoretical bounds (purple). The region within the black contour is allowed after the combination of all constraints.}
\label{fig:2}
\end{figure}

In Figure \ref{fig:2} I show the combination of all signal strengths transferred to the $\beta-\alpha$ vs. $m_H$ plane; it is independent of the heavy Higgs mass. I confront it with the regions disfavoured by all heavy Higgs searches, which mainly constrain $H$ masses below 1 TeV, and the scenarios excluded by the theoretical bounds, which for $m_H$ above 600 GeV push the 2HDM towards the alignment limit. While both, the heavy Higgs searches and the theoretical constraints hardly have an effect stronger than the signal strengths in this plane for type II, they are more relevant in type I, where the impact of the signal strengths is weaker. Finally, combining the LHC and theory constraints with the ones from flavour and Z-pole physics, one obtains the strips within the black contours. In both types, the global fit to all constraints only allows small deviations from the alignment limit; the type II $H$ mass additionally gets a lower bound of around 750 GeV if one simultaneously fits the lower bound on the charged Higgs mass from $b\to s \gamma$ transitions with the electroweak precision observables and the theory constraints (see also \cite{Cacchio:2016qyh,Chowdhury:2017xxx}).

\section{Summary}

After introducing the multi-purpose code HEPfit, I show its application to the 2HDM types I and II: I discuss the impact of the $h$ signal strengths and the searches for heavy Higgs bosons on the 2HDM parameter space. These measurements by the ATLAS and CMS collaborations yield strong bounds especially on the angle difference $\beta-\alpha$. With increasing data, it is being pushed more and more to the value of $\pi/2$, for which the 125 Higgs resembles the SM Higgs. For even more up-to-date fits, also to the two remaining types of $\mathbb{Z}_2$ symmetry I have not discussed here, I refer to \cite{Chowdhury:2017xxx}.

\acknowledgments

I thank Debtosh Chowdhury for helpful discussions.
This work was supported by the Spanish Government and ERDF funds from the European Commission (Grants No. FPA2014-53631-C2-1-P and SEV-2014-0398).

\bibliographystyle{JHEP}
\bibliography{2HDM_fits_with_HEPfit_bib}

\end{document}